\documentclass[twocolumn,showpacs,preprintnumbers,amsmath,amssymb]{revtex4}

\usepackage{graphicx}
\usepackage{dcolumn}
\usepackage{bm}
\usepackage{epstopdf}

\begin{document}

\preprint{APS/123-QED}

\title{Baryon Stopping as a new Probe
of Geometric Scaling}

\author{Yacine Mehtar-Tani}
\author{Georg Wolschin}%
\affiliation{%
 Institut f{\"ur} Theoretische 
Physik
der Universit{\"a}t Heidelberg, 
        Philosophenweg 16,  
        D-69120 Heidelberg, Germany
}%

\date{\today}

\begin{abstract}
We suggest to use net-baryon rapidity distributions in central relativistic heavy-ion 
collisions at SPS, RHIC and LHC energies in order to probe saturation physics.
Within the color glass condensate framework based on small-coupling QCD,
net-baryon rapidity distributions are shown to exhibit geometric scaling. 
In a comparison with RHIC data in Au + Au collisions at
$\sqrt {s_{NN}}$ = 62.4 GeV and 200 GeV the gradual approach to the
gluon saturation regime is investigated.
Predictions for net-baryon rapidity spectra and the mean rapidity loss in 
central Pb + Pb collisions at LHC energies of $\sqrt {s_{NN}}$ = 5.52 TeV are made.
\end{abstract}

\pacs{24.85.+p, 25.75.-q, 25.75.Dw, 12.38.Mh}
\maketitle
Baryon stopping in relativistic heavy-ion collisions  as a probe of QCD-matter at high parton density is of great current interest \cite{bea04,dal08,ita03,alb07}.  Theoretical QCD-based approaches usually focus on charged-hadron production. In the central rapidity region a reasonable understanding  has been achieved in the color glass condensate (CGC) framework \cite{ mcl94,Bal96,Jal97,Ian01} through inclusive gluon production \cite{nar05,alb07bis}. 
In this theory, due to the self-interaction of gluons, the number of gluons in the nuclear wave function increases with increasing energy and decreasing longitudinal momentum fraction $x$ carried by the parton.

Unitarity requires that the gluon density saturates below a characteristic momentum scale, the so-called saturation scale $Q_s$. In this regime gluons form a coherent state. Presently the evidence for the existence of this state of matter is, however, not yet clear. Due to the dependence of the saturation scale on rapidity and mass number, it has been proposed that saturation effects should be studied with heavy nuclei and large rapidities at RHIC energies and beyond.

\begin{figure}
\includegraphics[width=8cm]{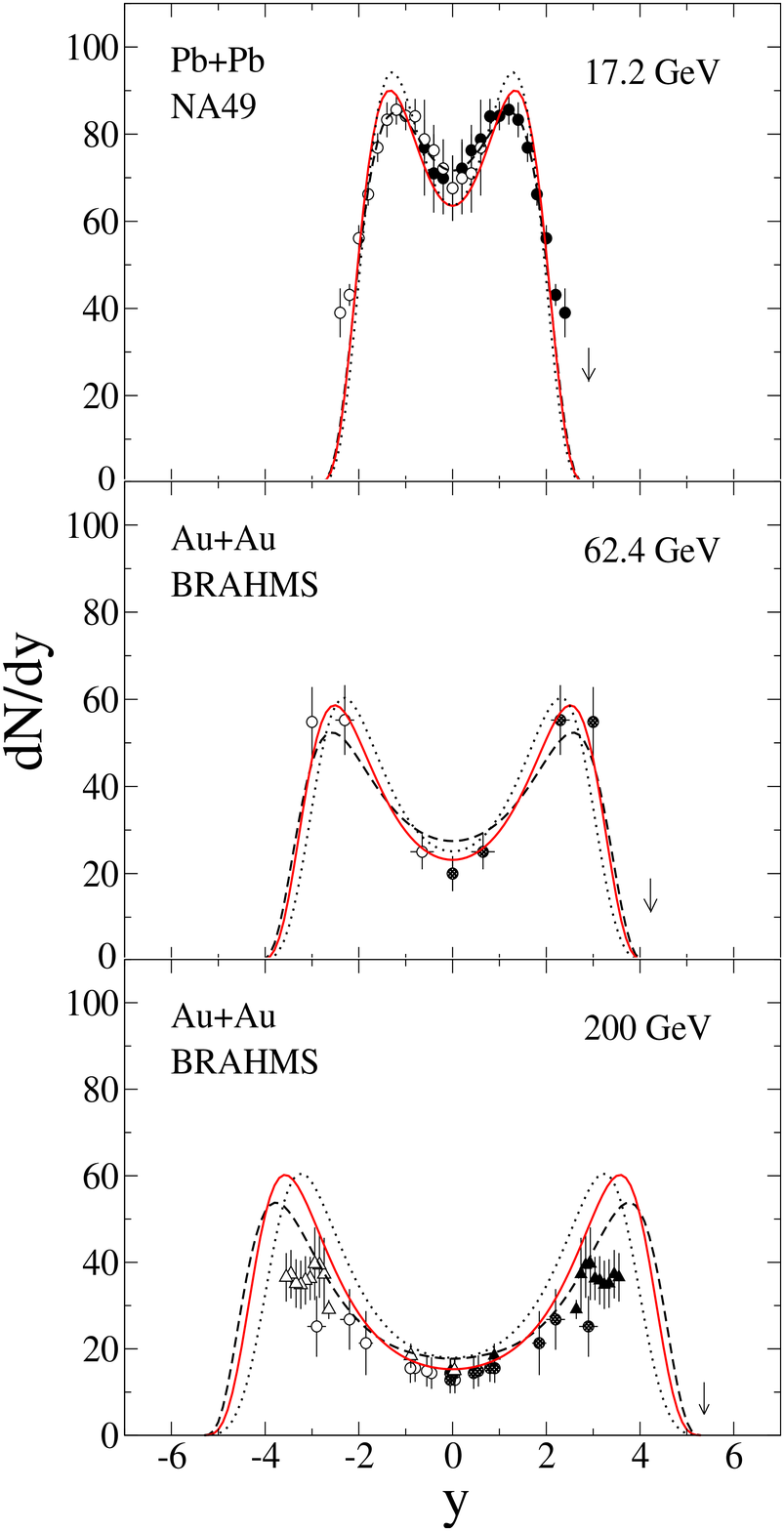}
\caption{\label{fig1} (color online). Rapidity distribution of net baryons
in central (0 -- 5\%) Pb + Pb collisions at SPS energies of $\sqrt {s_{NN}}$ = 17.2 GeV (top frame). The theoretical calculations are compared with
NA49 results that have been extrapolated from the net-proton data \cite{app99}. 
Dashed curves are for $\lambda = 0$ and $Q_0^2 = 0.08$ GeV$^2$,
solid curves are for $\lambda = 0.15$ and $Q_0^2 = 0.07$ GeV$^2$, and
dotted curves are for $\lambda = 0.3$ and $Q_0^2 = 0.06$ GeV$^2$.
At RHIC energies of $\sqrt{ s_{NN}}$ = 62.4 GeV (middle frame, 0 -- 10\%) and 200 GeV
for central Au + Au, our corresponding theoretical results are compared with BRAHMS net baryon data  (circles) \cite{bea04,dal08}. At 200 GeV, triangles are preliminary scaled BRAHMS net proton data points for 0 -- 10\% \cite{deb08}. }
\end{figure}

In this Letter we suggest to use the rapidity distribution of net baryons ($B-\bar{B}$) in central heavy-ion collisions as a testing ground for saturation physics, cf. Fig. \ref{fig1}. In $A + A$ collisions, two distinct and symmetric peaks with respect to rapidity $y$ occur at SPS energies \cite{app99} and beyond. The rapidity separation between the peaks increases with energy, and decreases with increasing mass number $A$ reflecting larger baryon stopping for heavier nuclei, as has been investigated phenomenologically in the relativistic diffusion model \cite{wols06}. 

The net-baryon number is essentially transported by valence quarks that probe the saturation regime in the target by multiple scatterings \cite{bas03}. During the collision the fast valence quarks in one nucleus scatter in the other nucleus by exchanging soft gluons, leading to their redistribution in rapidity space. 
We take advantage of the fact that the valence quark parton distribution is well known at large $x$, which corresponds to the forward  and backward rapidity region, to access the gluon distribution at small $x$ in the target nucleus. Therefore, this picture provides a clean probe of the unintegrated gluon distribution $\varphi(x,p_T)$ at small $x$ in the saturation regime. Here $p_T$ is the transverse momentum transfer.

We have two symmetric contributions, coming from the two beams. The contribution of the fragmentation of the valence quarks in the forward moving nucleus is given by the simple formula \cite{dum06} for the rapidity distribution of hadrons:



\begin{equation} \label{eq:yield}
\frac{dN}{dy}=\frac{C}{(2\pi)^2 }\int \frac{d^2p_T}{ p_T^2}\;x_{1}q_v(x_{1},Q_{f})\;\varphi\left(x_2,p_T\right),
\end{equation}
where  $x_1=p_T/\sqrt{s}\exp(y)$,  $x_2=p_T/\sqrt{s}\exp(-y)$ are the longitudinal momentum fractions carried, respectively, by the valence quark in the projectile and the soft gluon in the target.
The factorization scale is set equal to the transverse momentum, $Q_{f}\equiv p_T$. The contribution of valence quarks in the other beam nucleus is added incoherently by changing $y \to -y$. 
The gluon distribution is related to the forward dipole scattering amplitude ${\cal N}(x,r_T)$, for a quark dipole of transverse size $r_T$, through the Fourier transform
\begin{equation}\label{eq:FourierN}
\varphi(x,p_T)=2\pi p_T^2\int r_Tdr_T {\cal N}(x,r_T)J_0(r_Tp_T). 
\end{equation}
In the fragmentation region of the projectile the valence quark parton distribution function (PDF) is dominated by large values of $x_1$. 
We integrate out the fragmentation function such that the hadron rapidity distribution is proportional
to the parton distribution. 
 The overall constant $C$ depends on the nature of the produced hadron. 

\begin{figure}
\includegraphics[width=8.5cm]{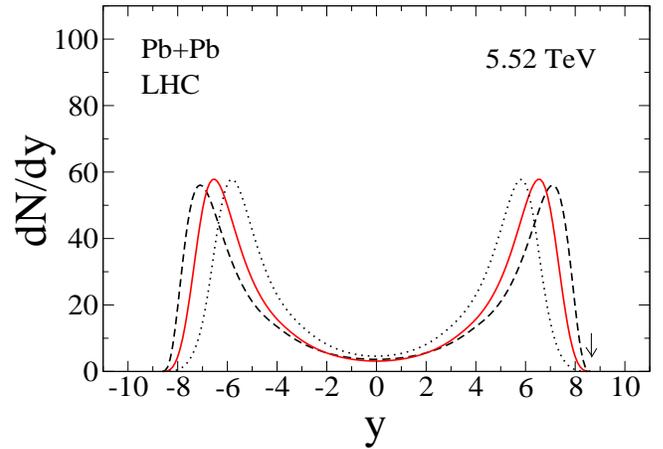}
\caption{\label{fig2} (color online). Rapidity distribution of net baryons
in central Pb + Pb collisions at LHC energies of $\sqrt {s_{NN}}$ = 5.52 TeV. Theoretical
distributions are shown 
for $\lambda = 0$ (dashed),  $\lambda = 0.15$ (solid), and $\lambda = 0.3$ (dotted curve),
with $Q_0^2$ - values as in Fig.1.}
\end{figure}
One important prediction of the color glass condensate theory is 
geometric scaling: the gluon distribution depends on $x$ and $p_T$ 
only through the scaling variable $p_T^2/Q_s^2(x)$, where $Q_s^2(x)=A^{1/3} Q_0^2\;x^{-\lambda}$, $A$ is the mass number and $Q_0$ sets the dimension. This has been confirmed experimentally 
at HERA \cite{sta01}. The fit value $\lambda = 0.2-0.3$ agrees with theoretical estimates based on next-to-leading order Balitskii-Fadin-Kuraev-Lipatov (BFKL) results \cite{lip76,tri03}. To show that the net-baryon distribution reflects the geometric scaling of the gluon distribution, we perform the following change of variables: 
\begin{equation}
x\equiv x_1,\;\;x_2\equiv x\;e^{-2y},\;\; p_T^2\equiv x^2 s\; e^{-2y}.
\end{equation}
Thus, we rewrite Eq. (\ref{eq:yield}) as 
\begin{equation}
\frac{dN}{dy}(\tau)=\frac{C}{2\pi}\int_0^1\frac{dx}{x}\;xq_v(x) \;\varphi(x^{2+\lambda} 
e^\tau),
\label{eq:GSyield}
\end{equation}
where $\tau=\ln (s/Q_0^2) - \ln A^{1/3} - 2(1+\lambda)\,y$ is the corresponding scaling variable. Hence, the 
net-baryon multiplicity in the peak region is only a function of a single 
scaling variable $\tau$, which relates the energy dependence to the rapidity and
mass number dependence. 
In the fragmentation region,
the valence quark distribution is only very weakly dependent on $Q_{f}$.

From the equation for the isolines, $\tau=\text{const}$, 
one gets the evolution of the position of the fragmentation peak in the forward region with respect to the 
variables of the problem
\begin{equation}\label{eq:peak}
y_{\text{peak}}=\frac{1}{1+\lambda}\left(y_{\text{beam}}-\ln 
A^{1/6}\right)+\text{const},
\end{equation}
where 
$y_{\text{beam}} =1/2\cdot \ln[(E + p_{L})/(E - p_{L})]\simeq \ln\sqrt s / m$
is the beam rapidity at beam energy $E$ and longitudinal momentum $p_{L}$ with the nucleon mass 
$m$. 

To take into account saturation effects in the target we choose the Golec-Biernat-W\"usthoff model \cite{GBW98} for the  forward dipole scattering amplitude $\cal N$, leading to (cf. Eq. (\ref{eq:FourierN}) and \cite{dum06})
\begin{equation}
\varphi(x,p_T)=4\pi\frac{p_T^2}{Q_s^2(x)}\exp\left(-\frac{p_T^2}{Q_s^2(x)}\right).
\end{equation}
The valence quark parton distribution of the nucleus is taken to be equal to the valence quark PDF in a nucleon times the number of participants in the nucleus. 
We are focusing here on the forward rapidity region, and interpolate to mid-rapidity where small-$x$ quarks are dominant, by matching the leading-order distributions \cite{mrst01} and the Regge trajectory, $xq_v\propto x^{0.5}$, at $x =0 .01$ \cite{ita03}. 

To account for large-$x$ effects in the gluon distribution, we multiply the distribution function by $(1-x_2)^4$ \cite{alb07bis}. Mass effects are considered through the replacement $p_T\rightarrow \sqrt{p_T^2+m^2}$. 

Our results for net-baryon rapidity distributions in central Pb + Pb and Au + Au collisions are shown in Fig.   \ref{fig1}.
Dashed curves are for $\lambda = 0$,
solid curves for $\lambda = 0.15$,
and dotted curves for 
 $\lambda = 0.3$,
 with the corresponding $Q_0^2$-values fixed at SPS, as given in the caption.
 
We compare with SPS NA49 Pb + Pb results at $\sqrt{s_{NN}} = 17.2$ GeV \cite{app99}, and BRAHMS Au + Au data at $62.4$ GeV and $200$ GeV \cite{bea04,dal08,deb08}. We obtain the number of baryon participants  at SPS energy in the full rapidity range from a double-gaussian fit of the NA49 data
for  $0 - 5\%$ central  Pb + Pb collisions as $N_B = 380$. The normalization in our model calculation is
$12\%$ lower than this value since we do not account for the baryons near $y_{\text{beam}}$ in the tails. We maintain this correction at RHIC energies where the tails are in the unmeasured region. 

For Au + Au at RHIC energies, we take Glauber results for the number of participants: At 62.4 GeV $N_B = 314\pm 8 $ for centrality $0 - 10\%$ \cite{dal08} , and at 200 GeV $N_B = 357\pm 8 $ for $ 0 - 5\%$ \cite{bea04}. The comparison with the SPS and RHIC data slightly favors $\lambda \leq 0.15$ and hence, the asymptotic regime with  $\lambda \simeq 0.3$  is not yet reached at RHIC. 

Our prediction for central Pb + Pb at $5.52$ TeV LHC energies is shown in Fig. \ref{fig2} for $\lambda = 0, 0.15$, and 0.3. At LHC energies the mid-rapidity region is almost baryon free, we obtain $dN/dy (y=0)\simeq 4$ for net baryons. The
position of the fragmentation peak is very sensitive to the value of $\lambda$, with a difference of about 1.5 units of rapidity  between the $\lambda = 0$ and 0.3 cases. It is possible that
the full scaling regime with $\lambda$ approaching 0.3 can be reached at or beyond LHC energies, but presently none of the LHC-experiments is capable of measuring identified protons or neutrons from central Pb + Pb collisions in the region of the fragmentation peaks. This would be a relevant proposal for future extensions of the detector capabilities at LHC. 

Physically, the two peaks represent the result of the scattering of the fast moving projectile valence quarks in the target, they are deflected, their distribution broadens and carries information about the gluon distribution in the target. This is in analogy to x-rays that are deflected by a crystal and carry information about its structure.

With increasing energy the peaks move apart, the solutions behave like travelling waves in rapidity space \cite{mun03}, which can be probed experimentally at distinct values of the beam energy, or the corresponding beam rapidity. We have derived the peak position as a function of the
beam rapidity as $y_{\text{peak}} = v\, y_{ \text{beam}} + \text{const}$ 
with the peak velocity
 $v = 1/(1+\lambda)$, cf. Eq. (\ref{eq:peak}). The position of the peak in rapidity space as a function of the beam energy can in principle be determined experimentally, or at least estimated (RHIC). Theoretically, its evolution with energy provides a measure of the saturation scale exponent $\lambda$. Hence, a precise determination of the net-proton fragmentation peak position as a function of beam energy would provide detailed information about the gluon saturation scale.

\begin{figure}
\includegraphics[width=7.5cm]{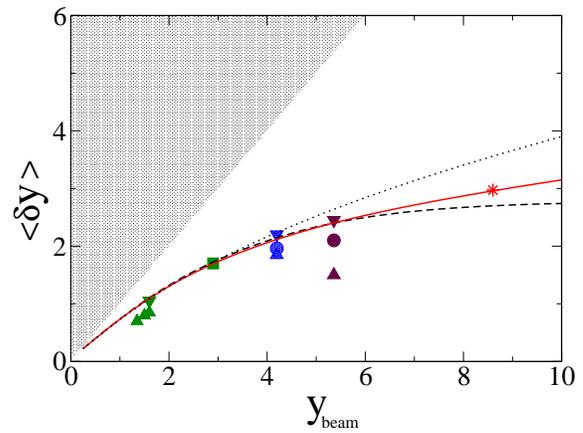}
\caption{\label{fig3} (color online). The mean rapidity loss $\langle \delta y\rangle$ as obtained from
our theoretical results is plotted as a function of beam rapidity $y_{\text{beam}}$, solid curve.
The star at  $y_{\text{beam}}$ = 8.68  is our prediction for central Pb + Pb at LHC-energies of
$\sqrt {s_{NN}}$ = 5.52 TeV with $\lambda = 0.15$, the dashed curve is for   $\lambda = 0$, the dotted curve is for  $\lambda = 0.3$, with $Q_0^2$ values as in Fig.1.
Analysis results from AGS Au + Au data (E917, E802/E866, triangles) \cite{vi95},
SPS Pb + Pb data (NA49, square) \cite{app99}, RHIC Au + Au data (BRAHMS, dots, with triangles as lower and upper limits) \cite{bea04,dal08} are compared with the calculations.}
\end{figure}

In Fig. \ref{fig3}, we show our numerical results for the mean rapidity loss $\langle \delta y\rangle=y_{\text{beam}}-\langle  y\rangle$. At low energies they agree with the experimental values of baryon stopping that have been obtained at AGS and SPS energies \cite{vi95,app99}
 irrespective of the value of $\lambda$. Here we have considered the effect of the missing particles in the  tails 
as described above for the rapidity distributions, placing them halfway between the mean rapidity, and the beam value. 
At  RHIC energies of 62.4 GeV and 200 GeV, the mean rapidity loss depends on $\lambda$, and the $\lambda = 0.3$ result (dotted curve) is beyond the upper limit given by BRAHMS \cite{bea04,dal08}, whereas $\lambda \leq 0.15$ is consistent with the upper limit of the data. 
Consequently, up to the highest RHIC energies the expected scaling regime with $\lambda \simeq 0.3$ \cite{sta01} is not yet fully reached, in accordance with  \cite{alb07bis,hir04}.

Our result emphasizes the importance of a detailed measurement at LHC energies to allow more definite conclusions about the value of $\lambda$, which would then be determined by the slope of the mean rapidity loss at high
beam rapidity above RHIC (solid curve in Fig. 3).

Assuming that the mean rapidity evolves similarly to the peak position, $\langle y\rangle\equiv y_{\text{peak}} + $const., the linear increase of the mean rapidity loss at large energies corresponding to beam rapidities
$y_{\text{beam}} >  5$ is given by 
\begin{equation}
\langle \delta y\rangle = \frac{\lambda}{1+\lambda}y_{\text{beam}}+\text{const}.
\end{equation}\\
Hence, the mean rapidity loss that accompanies the energy loss in the course of the slow-down of baryons provides at large beam rapidities $y_{\text{beam}} >  5$  a measure for $\lambda$ and thus, a test of saturation physics. The case $\lambda=0$, or equivalently $Q_s$ constant, leads to a saturation of the mean rapidity loss at high energies, and correspondingly at large beam rapidities.

In the peak region, the average $x$ in the projectile is $x\simeq 0.2-0.3$, which corresponds to the average momen-\\
tum fraction carried by a valence quark. In the target,  $x = (0.2-0.3)\,e^{-2y_{\text{peak}}}$, it decreases with increasing energy. In this kinematic regime we have a natural intrinsic hard momentum, the saturation scale $Q_s$. This justifies the use of small-coupling techniques in QCD for calculating integrated yields \cite{dum03}. The effects of the medium are expected to be small at forward rapidity since the fast moving valence quarks escape the interaction zone quickly. A detailed measurement of the peak region would then enable us to reconstruct the gluon distribution from Eq. (\ref{eq:yield}). 

To summarize, we have presented a saturation model for net-baryon distributions to investigate the
gradual approach to the gluon saturation regime at RHIC energies and beyond. In a comparison with BRAHMS net-baryon results for central Au + Au collisions at 62.4 and 200 GeV we have determined a saturation-scale exponent $\lambda \leq 0.15$ and hence, the full scaling regime is not yet reached at RHIC. This result is in agreement with studies of particle production that point out a slower growth of the saturation scale at RHIC energies than the HERA estimate of $\lambda \simeq 0.3$ suggests \cite{alb07bis,hir04}.

In particular, we have shown that the peak position in net-proton rapidity distributions of centrally colliding heavy ions at  ultra-relativistic energies obeys a scaling law involving the atomic mass and the beam energy. Our result for the mean rapidity loss in $\sqrt{ s_{NN}}$ = 62.4 GeV and 200 GeV Au + Au is for $\lambda \leq 0.15$ consistent with the upper limit of the corresponding BRAHMS experiments. We emphasize the importance of a detailed analysis at LHC energies.

One of the authors (Y. M.-T.) acknowledges
critical comments by Jamal Jalilian-Marian and Mark Strikman.
We are grateful to the BRAHMS
collaboration for  
their data.
This work has been supported
by the Deutsche Forschungsgemeinschaft under Grant No. STA 509/1-1.

\end{document}